\documentclass[sn-aps,iicol]{sn-jnl}

\usepackage{Tomspeak}
\newcommand{\etal}{{\it et al. }}

\newcommand{\dmax}{d_\text{max}}

\usepackage{bm}
\usepackage{hyperref}
\usepackage{graphicx}%
\usepackage{multirow}%
\usepackage{amsmath,amssymb,amsfonts}%
\usepackage{mathrsfs}%
\usepackage[title]{appendix}%
\usepackage{xcolor}%
\usepackage{textcomp}%
\usepackage{manyfoot}%
\usepackage{booktabs}%
\usepackage{algorithm}%
\usepackage{algorithmicx}%
\usepackage{algpseudocode}%
\usepackage{listings}%

\begin{document}

\title{Granular convergence as an iterated local map}

\author[1]{\fnm{Anna} \sur{Movsheva}}\email{amovsheva@uchicago.edu}

\author*[1]{\fnm{Thomas A.} \sur{Witten}}\email{t-witten@uchicago.edu}

\equalcont{These authors contributed equally to this work.}

\affil[1]{\orgdiv{James Franck Institute}, \orgname{University of Chicago}, \orgaddress{\street{929 E. 57th Street}, \city{Chicago}, \postcode{60637}, \state{Illinois}, \country{USA}}}


\abstract{Granular convergence is a property of a granular pack as it is repeatedly sheared in a cyclic, quasistatic fashion, as the packing configuration changes via discrete events. Under suitable conditions the set of microscopic configurations encountered {\em converges} to a periodic sequence after sufficient shear cycles. Prior work modeled this evolution as the iteration of a pre-determined, random map from a set of discrete configurations into itself.  Iterating such a map from a random starting point leads to similar periodic repetition.  This work explores the effect of restricting the randomness of such maps in order to account for the local nature of the discrete events.  The number of cycles needed for convergence shows similar statistical behavior to that of numerical granular experiments.  The number of cycles in a repeating period behaves only qualitatively like these granular studies.  }




\maketitle

\section{Introduction}\label{sec:Introduction}
In the 1980's soft matter science was largely confined to molecular structures such as polymers.  Polymers showed that qualitative rules of construction of a many-body system could yield powerful predictions about their behavior.  Already in 1982 in the Pincus-Edwards workshop in Santa Barbara\cite{Edwards:1985dq} this same power began to be applied to assemblies of macroscopic objects, such as granular packs\cite{Edwards:1982vf}.

\begin{figure}
    \centering
    \includegraphics[width=.22\textwidth]{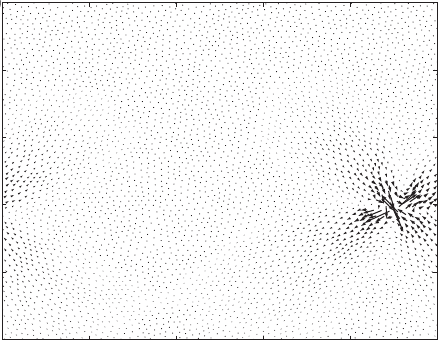}
    \includegraphics[width=.23\textwidth]{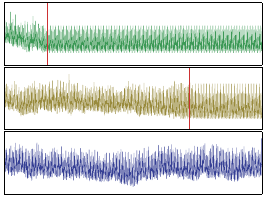}

    \caption{ Slip events and cycles in a sheared bead pack. Left:  configuration of simulated pack, after Ref. \cite{Regev:2015fk} (courtesy of the authors) showing  displacements during a slip event. Right: graphs of potential energy \vs cycle number showing initial secular motion followed by a periodic motion for three different shear amplitudes $\Gamma$. The lowest amplitude in the top graph shows periodic behavior after many cycles; a graph with a slightly larger amplitude requires a longer time to converge.  Small additional amplitude yields variability that never becomes periodic. }
\label{fig:simulation_cycles}
\end{figure}

We focus here on a remarkable form of self-organization in a granular pack that we call ``convergence", revealed in atomistic computer simulations\cite{Fiocco:2014uq,Mungan:2019bc,Mungan:2021we,Sastry:2021vk,Regev:2013fk,Regev:2015fk,Lavrentovich:2017rz}.  One subjects a disordered, mechanically solid  packing of spheres to repeated cycles of gradual shear deformation while tracking the precise  positions of all the spheres---the  ``configuration"---through repeated cycles of alternating back-and-forth shearing, during which many spheres change their relative positions.  Initially, the configurations change with each succeeding  cycle.  But after a number of cycles, the complicated sequence of rearrangements converges to an exact repetition, as evidenced in Fig \ref{fig:simulation_cycles} (right).  The configurations encountered during the forward and the reverse shearing are different.    A cycle may generate dozens of configurational shifts, selected from the vast number of possible configurations.  Any variations in the shearing cycles yield convergence to a different set of configurations.  Though the sequence may repeat after a single shear cycle, it may also reach a state requiring two or more shear cycles in order to return to its initial state.  

In this paper we aim for an improved understanding of this convergence process using the concept of discrete maps defined  in Ref. \cite{Mungan:2019fk}.  First, we review the rationale for the discrete map picture.  

Under cyclic annealing, one imposes shear strains \eg by translating the upper confining boundary of the pack alternately left and right, thus cycling the shear strain $\gamma$ between bounds $-\Gamma$ and $\Gamma$.   The particles move quasistatically, following a local minimum in the potential energy as the shear changes.  This changing shear strain produces abrupt slip events such as that pictured in Fig. \ref{fig:simulation_cycles}a,  when the particles have reached an unstable configuration.  

Each slip event amounts to a discrete change in the positional configuration $c$ of the pack.  Given a configuration $c_i$ just after a slip event, there is a continuous  change of $c$  determined by the increasing shear, leading to an unstable state and a slip event to a new configuration denoted $c_j$.   Following Ref. \cite{Mungan:2019fk}, we postulate that this $c_j$ is a fixed function denoted $\mu(c_i)$.  That is, whenever some slip event puts the pack into some configuration $c_i$, the next slip event caused by increasing shear must be the unique configuration $c_j$ dictated by the deterministic change from $c_i$ leading through the subsequent slip event \footnote{These $c_i$ are an alternative to the ``mesostates" defined in Refs. \cite{Fiocco:2014uq,Mungan:2019bc,Mungan:2021we,Sastry:2021vk,Regev:2013fk,Regev:2015fk},  in terms of particular ranges shear amplitude within which  no slip events occur. }.
A separate function $\nu(c_i)$ governs changes under decreasing shear\footnote{In Ref. \cite{Mungan:2019fk} the functions $\mu$ and $\nu$ are denoted $U$ and $D$, respectively.}. The result of an entire cycle of shear can thus be determined by iterating the functions $\mu$ and $\nu$, as pictured in Fig. \ref{fig:munu+M}. We denote the configuration obtained after an entire shear cycle by $M(c_i)$.


In  what follows we extend the deterministic map picture sketched above.  We ask how it should be modified to take into account two features of granular annealing: extensivity with system size and the locality of the slip events.    Section \ref{sec:extensivity} addresses how local maps can be compatible with the extensive behavior of the pack as a whole.   In Sections \ref{sec:Maps} and \ref{sec:Minimal} we define a crude implementation of local maps.   
In Sec. \ref{sec:scaling} we report the scaling behavior of two characteristic features: the number of cycles needed for convergence, and the number of cycles making up the final multicycle repetition.   
\section{Extensivity and Locality}
\label{sec:extensivity}
As noted above, granular packs in the regime of interest are  presumed extensive, so the number of elementary slip events in a given interval of shear must be proportional to the number of particles in the pack.  Thus the map for two successive elementary slip events is not local in the sense outlined above.  To recover a sense of locality one must introduce a notion of a correlation volume for a given  shear amplitude $\Gamma$.  We are interested in full cycle maps $M(c_i)$ that  eventually repeat in one or more shear cycles, \ie $M(c_i) = c_i$ or $M(M(c_i)) = c_i$, etc.   

 As shown in Fig. \ref{fig:simulation_cycles} (left), the local rearrangement of particles in a slip event produces a displacement field elsewhere.  Still, the slip event may be regarded as local.  The volume $V_o$ affected is  presumed independent of system size and comparable to the volume of a few particles.  We shall neglect any causal effect between this slip region and all others located outside of its volume during a shear cycle.  

\begin{figure}
    \centering
    \includegraphics[width=.45\textwidth]{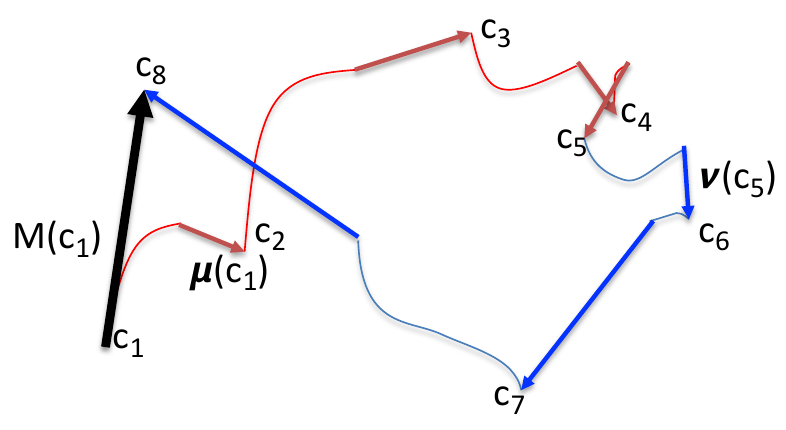}
    \caption{Comparison of incremental maps $\mu(c)$ and $\nu(c)$ \vs cycle maps $M(c)$, as used in Ref. \cite{Mungan:2019fk}. The lines show a possible motion of some particle within the region as the shear increases from $-\Gamma$ to $+\Gamma$ and back to $-\Gamma$.  The motion begins just after a slip event to configuration $c_1$.  The continuous, deterministic motion from the $c_1$  event, as shear increases, is shown by a thin line terminating with a second slip event, marked by a light-colored (red) arrow, and landing at $c_2 \definedas \mu(c_1)$.   Further increases in shear lead to $c_3, c_4$, and $c_5 = \mu(\mu(\mu(\mu(c_1))))$ . The remaining steps in the cycle, to configurations $c_6, c_7$ and $c_8$ result from decreasing shear, controlled by $\nu(c)$, with slip events indicated by dark-colored (blue) arrows. The cycle map $M(c_1)\definedas c_8$ is shown by a heavy black arrow.}
    \label{fig:munu+M}
\end{figure}

 We cannot neglect the possibility that a second slip event may occur {\em inside} the volume $V_0$ during a cycle, thus disturbing its determinism.  Now, this probability depends on the shear amplitude $\Gamma$.  It can be expressed in terms of the (intensive) density $\rho(\Gamma)$ of elementary events in a shear cycle.  This density becomes larger as $\Gamma$ increases.  We shall consider the regime where $\rho(\Gamma) \muchlessthan  1/V_o$, so that the $V_o$ region evolves  virtually independently of other regions; that is, the region is ``autonomous".  Its changes of state are dictated by elementary slip-event maps $\mu(c)$ and $\nu(c)$  for configurations $c$ within the region.  By iterating  either of these maps, one may infer the effect of a given finite change of the imposed shear, including an entire shear cycle, as pictured in Fig. \ref{fig:munu+M}.  
The result of $k$ shear cycles is then given by the $k$-th iterate of $M$ from the initial state $c$, \ie  $M(M(M(...M(c)....)))$.

In the numerical studies of Refs. \cite{Lavrentovich:2017rz,Mungan:2019bc,Regev:2013fk,Regev:2015fk}, all slip events occurring throughout the simulation are recorded.  In our interpretation, this includes many autonomous $V_o$ regions.  Each cycle of shear alters the state $c$ of a region via its own map $M(c)$.  The system as a whole only reaches its final repeating state when all of its autonomous regions have reached theirs. The number of shear cycles needed for convergence is thus governed by that autonomous region which needs the most shear cycles in order to converge to its final periodic behavior.  This feature will be important in the discussion below.

\section{Maps and graphs}\label{sec:Maps}
 The scheme above can be used to determine the fate of any system of discrete elements $\{c_i\}$ after many iterations of a map function such as $M(c)$.  A number of varieties were explored in Ref. \cite{Mungan:2019fk}.  It is convenient to represent a given map by its directed graph, in which the configurations $c$ are the nodes and the image of a given configuration and the map from $c$ to its image $c' \definedas M(c)$ is represented by an arrow from $c$ to $c'$.  Thus the graph has exactly one directed edge from each of its nodes.  
\begin{figure}
    \centering
\includegraphics[width=.45\textwidth]{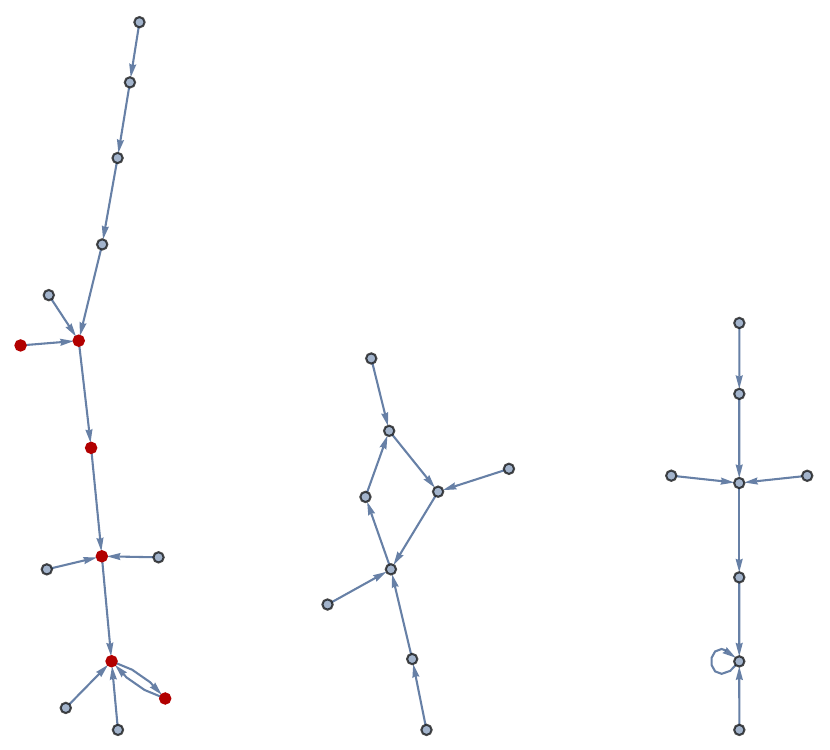}
    \caption{Graph of a random map on 32 nodes. The node positions are arranged to show connectivity clearly. Small arrows indicate the map from each node to its image. This map has three components.  The path from the leaf node at left to its successive images is shown in color (red).  This path terminates in a repetition of two nodes.  The middle component has a final repetition of four nodes. The right-hand component has a fixed point, shown as an arrow from a node to itself.}  
    \label{fig:RandomMap}
\end{figure}

The properties of generic graphs are described in Ref. \cite{Flajolet:1990fk}. Several of these general properties will be useful below.  First, a general graph consists of disjoint, unconnected subgraphs called {\em components}, as shown in Fig. \ref{fig:RandomMap}.  Within each component there are generally nodes that are not images of any others;  their only connection is from the node to its image.  These nodes are the {\em leaves} of the component.  All the nodes in a component can be traversed by iterating from each of its leaves.  During iteration from a leaf, each new image may be a node not previously encountered.  However, at some stage, a node is reached that {\em was} encountered.  Further iterations repeat the sequence of nodes already encountered, thus forming a {\em loop}.  The sequence of nodes from a given node to the loop is called the {\em tail} of that node.
Any two leaves with different loops are necessarily disjoint and must belong to separate components. Thus every component has exactly one loop, which may be a single node.
\begin{figure}
    \centering
    \includegraphics[width=.45\textwidth]{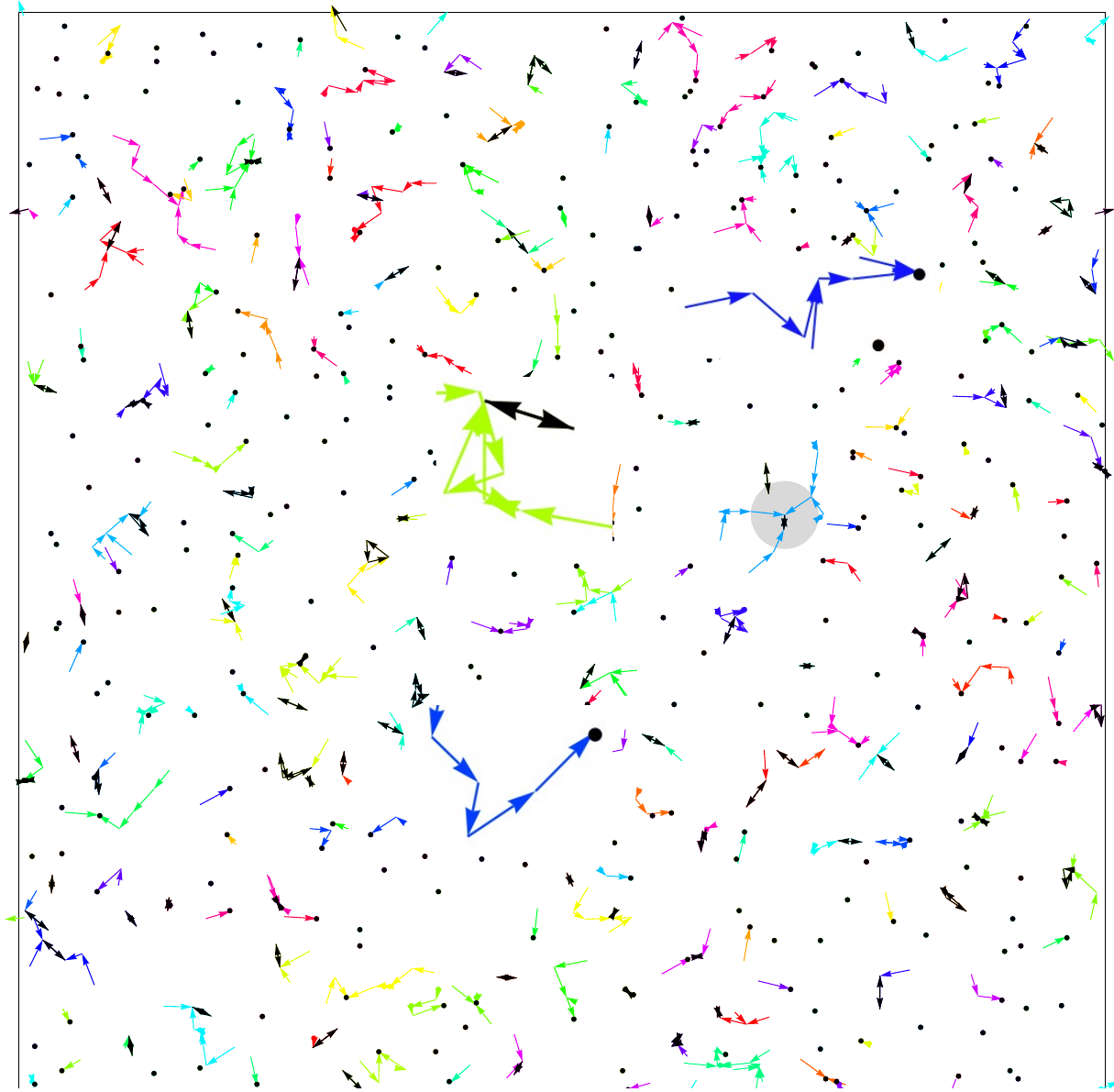}
    \includegraphics[width=.45\textwidth]{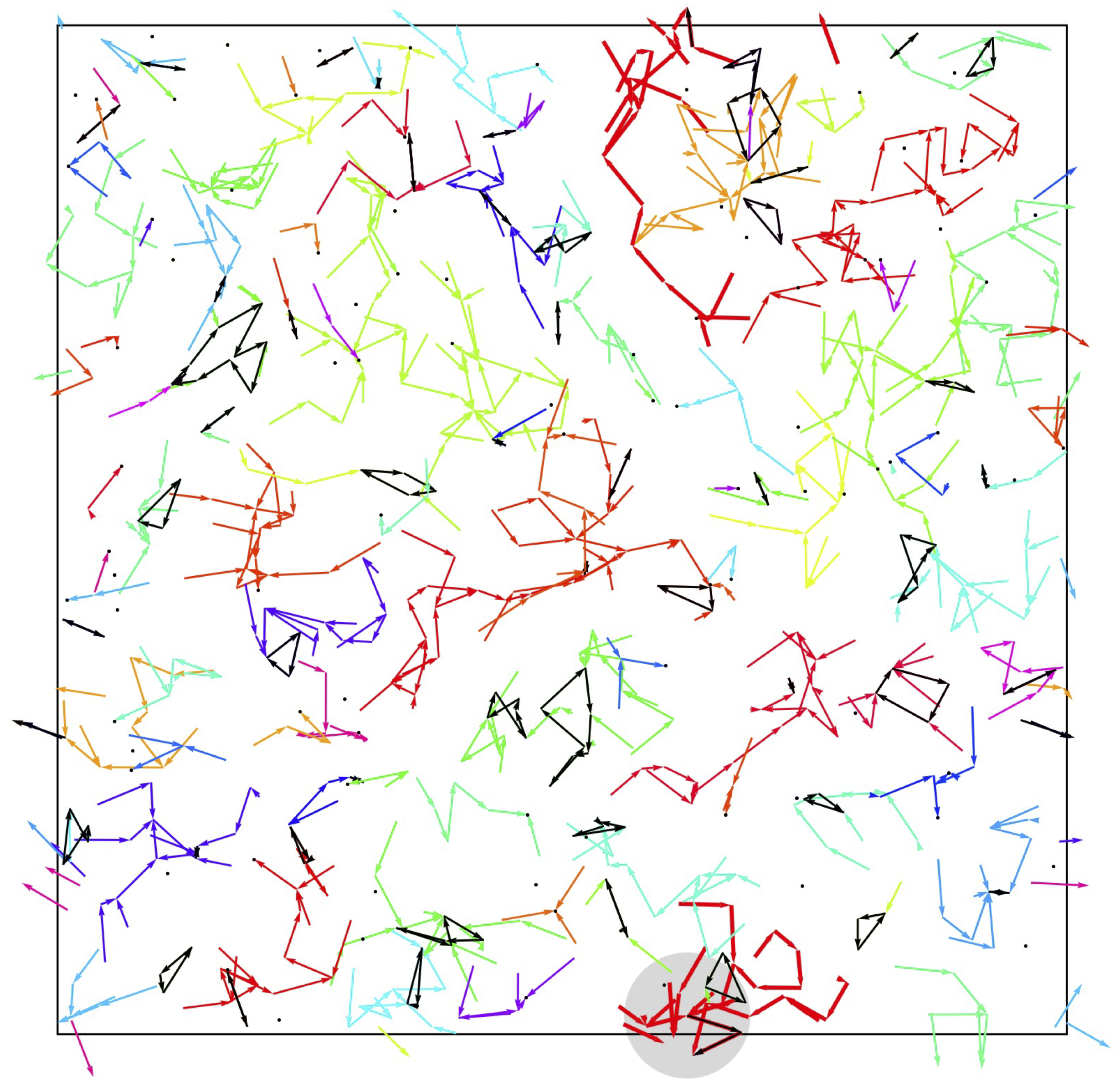}
    \caption{Example graphs of local maps. The black dots are nodes. The 1024 nodes are placed randomly in a square with periodic boundaries.   Colored arrows indicate the image node for every node.  A different color is chosen randomly for every component.  Black arrows indicate the loop for each component.  Single node loops are not shown.  Gray disk indicates the maximum image distance $\dmax$.  Top: Small $\dmax$. Bottom: Larger $\dmax$.}
    \label{fig:oldmaps}
\end{figure}
\section{Minimal local map}\label{sec:Minimal}
\begin{figure*}
\centering
\includegraphics[width=0.24\textwidth]{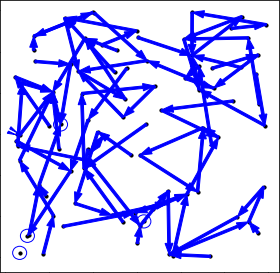}
\includegraphics[width=0.24\textwidth]{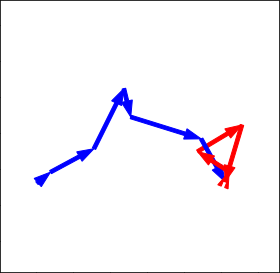}
\includegraphics[width=0.24\textwidth]{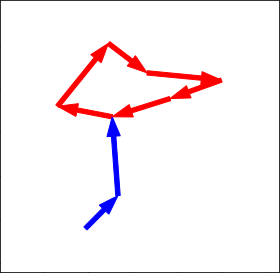}
\includegraphics[width=0.24\textwidth]{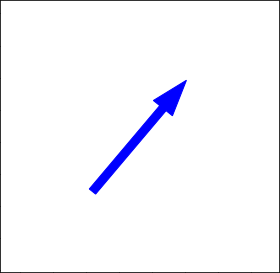}\\
\includegraphics[width=0.24\textwidth]{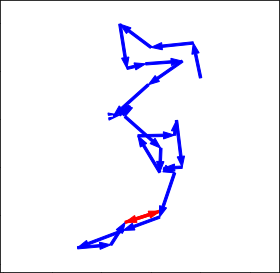}
\includegraphics[width=0.24\textwidth]{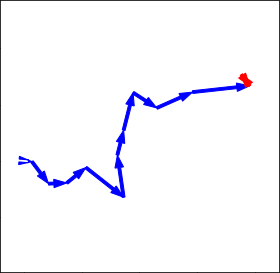}
\includegraphics[width=0.24\textwidth]{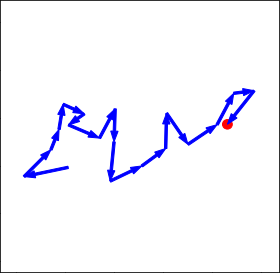}
\includegraphics[width=0.24\textwidth]{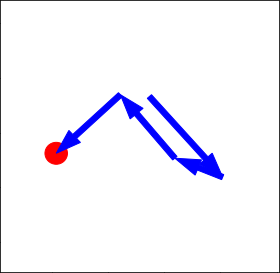}\\
\includegraphics[width=0.24\textwidth]{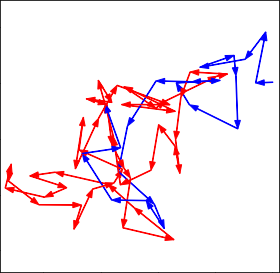}
\includegraphics[width=0.24\textwidth]{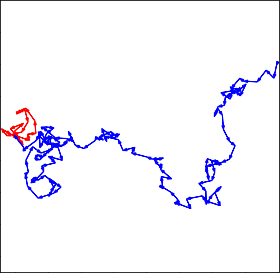}
\includegraphics[width=0.24\textwidth]{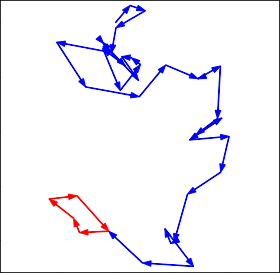}
\includegraphics[width=0.24\textwidth]{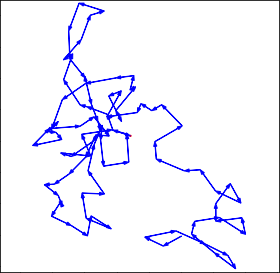}

\caption{Sampled iterates of randomly-selected nodes with local maps \cf Sec. \ref{sec:Minimal} . (Top left) Graph for a full map with $m = 32.169$. A blue circle around a point means the point has a directed edge to itself. Each point only has one edge leaving it. (Two top rows) Sampled  map iterates of random initial nodes with $m=32.169$"
. Light-colored (red) arrows indicate the terminating loop. (Bottom row) Sample graphs for $m = 804.248$.}
\label{fig:tailsAndLoops}
\end{figure*}

 As shown in Ref. \cite{Mungan:2019fk}, systems governed by discrete maps  exhibit the qualitative property of convergence seen in sheared granular packs.  However, these general maps give no predictive information about this convergence.  They take no account of the extensivity in system size expected for real granular packs. Further, they do not account for the locality of the events that constitute the evolution.

Here we define a further type of map that aims to address these deficiencies.  That is, for each configuration $c$, the range of possible images $c'$ is strongly limited.  The notion of locality requires a notion of distance; accordingly, we view our nodes, \ie the configurations $c$, as occupying points in a space.  For most of our work, we have used a two-dimensional space. This hypothetical dimension should not be confused with the dimensionality of the pack, \cf Sec. \ref{sec:implementingLocality}.  

We define our maps such that the distance from a node to its image must be less than a limiting distance denoted $\dmax$. A simple implementation of such a map is to place the node points at random in the space and then choose the image of each point by selecting the available points within distance $\dmax$ at random.  By making this selection for every node point in the space, we have defined a local map, as shown in Fig. \ref{fig:oldmaps}. Given an explicit map graph, one may easily determine successive images by following the arrows of the graph.  As with the generic graphs described above, these graphs have components, tails, loops, and leaves.  

The structure of these graphs depends on the limiting distance $\dmax$.  The essential effect of $\dmax$ is to alter the number $m$ of potential images.  The average $m$ is evidently the density of points $\rho$ times the area within $\dmax$:  
\begin{equation}
    m \definedas \pi \rho ~\dmax^2 .
\end{equation}  Henceforth we will use this ``range parameter" $m$ as our measure of locality.  

\subsection{Implementation} 
 If an autonomous region of a pack were to change its configuration over a shear cycle according to the local maps of Sec. \ref{sec:Minimal}, it would take one step along the graph in every shear cycle, until it reached the terminating loop.  The number of steps to reach the terminating loop is then the tail length of the graph from a random starting node.  Likewise, once the repeating behavior has begun, the number of cycles required for a repetition of its states is the loop length of the graph.  Accordingly, in the analysis below we focus on the statistics of the loop and tail lengths of our maps, with an eye to comparing these with packing simulations.  
\subsubsection{Tail and loop statistics}
The statistics of tails and loops are especially easy to sample for the local maps defined above.  We wish to know the distribution of tail and loop lengths for graphs containing randomly-chosen nodes.

We construct the maps using the object constructs in the Python language.  Our typical simulations use a density of $2^{10}$ \ie 1024. Each node is an object containing its x-y position and a pointer to its eventual image node.  

Once the node positions have been assigned, we may measure tail and loop lengths without constructing the entire map.  
We first make a list of the node objects with randomly-chosen x-y point coordinates. Next, for each node, we compile a neighbor list of the nodes within distance $\dmax$ of that node. This list includes the node itself. Then for each node, we choose an image node from its neighbor list at random. We set the pointer in the node to its image node. To construct a sample of tail lengths, we choose a node at random and we follow its pointer to the next node again and again, marking each node as “visited”. As we follow the pointers to the next node and the next node, we count the number of nodes we have visited. This counting and traversing of nodes ends once we come across a node that has already been marked as “visited”.
Thus the prior node was the beginning of a loop, and all further iterations repeatedly traverse this loop.  The nodes we have traversed so far are the tail-plus-loop for the given starting node. The tail is the sequence from the starting node to the  node that began the loop.  The remaining nodes in the tail-plus-loop form the loop. 
We then store the lengths of the tail  and loop as one sample of the desired distribution.  We repeat this process to obtain the entire distribution.
Fig. \ref{fig:tailsAndLoops} shows typical samples.

\subsubsection{Component statistics} \label{sec:componentStatistics}

To find a component of a graph, we must identify all the nodes of each component. For our needs, it suffices to determine the average number of nodes in a component weighted by per-node sampling.  Given an arbitrary node $c_i$, we can determine the number of nodes $C_i$ in its component by generalizing the procedure above for identifying the tail and loop for a given node.

Before connecting the nodes, all the nodes start off being leaves. Accordingly, we may set a logical ``is\_leaf" variable in each node indicating that it is a leaf. As each node is being connected to another node, the ``is\_leaf" variable of its image  node is cleared.  Once the full map is constructed, its full set of leaves is established.

Unique components are fully disjoint so they do not share any nodes; in other words, each node will belong to only one component. To determine the nodes of a component, we begin by assigning a unique index to each leaf.  We then iterate the map from each leaf, labeling each node encountered with the leaf index.  When such a node already has a leaf index (because it is the descendent of a previously-iterated leaf), the current leaf index is changed to match the previously assigned one.  Its descendants are also relabeled. Any two leaves that belong to the same component will eventually acquire a common index, and so will the descendants of these leaves. 

When all the leaves of the map have been processed in this way, all nodes in a given component will have the same index, and each component will have a unique index\footnote{Any component consisting entirely of a loop will not be counted in this way, since such a component has no leaves. We treat these in Appendix \ref{app:noLeaves} } To find the size $C_i$ of a particular component, we find and count all the nodes with its index. The average of all the $C_i$ sampled in this way is denoted $C$.

In a map with many configurations, it is not practical or necessary to determine all the components.  Instead, we create many maps and study a small class of components with nodes near the center of the sample box. For small $m$ these components thus found are concentrated near this initial node, with no nodes near the boundaries.  

For larger $m$ such that the identified nodes extend beyond the middle quarter of the box, it is possible that the component would change if the box were larger. When such components are encountered,  we extend the box in all directions filling it with the same density of randomly placed points. We then construct the map images for these new nodes and determine the leaves and component indices for these nodes as above.  We continue this process until the new nodes have no effect on the component under study.  We provide further details in Appendix \ref{app:largeComponents}. 

\section{Scaling of convergence properties}\label{sec:scaling}
In this section we report the effects of our range parameter $m$; this is our proxy for the shear amplitude $\Gamma$. This proxy is plausible, since increasing the shear amplitude increases the number of ways an instability can occur in an autonomous region, and thence the number of nodes needed to represent it.  Average tail and loop lengths increase with $m$ corresponding to the increased number of shear cycles for annealing. For fixed $m$ these lengths remain finite for arbitrarily large map sizes.  One readily obtains tail lengths that are numerically similar to the number of cycles to convergence observed in particle-scale   simulations\cite{Regev:2015fk,Lavrentovich:2017rz}. 

Another aspect to be compared to simulations is the ratio of loop to tail lengths.  For random maps, these are equal up to a factor of order unity\cite{Flajolet:1990fk}.   By contrast, in the granular simulations the tail to loop length ratio appears to vanish in the limit of large loop length.

\begin{figure}
    \centering
    \includegraphics[width = 0.4\textwidth]{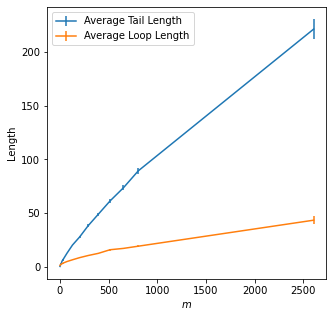}
    \caption{Average loop and tail size of graphs, for a given range parameter $m$ (the average number of candidate image points for each point). The underlying parameter is $\dmax$. }
    \label{fig:loop_and_tail_vs_n}
\end{figure}
\begin{figure}
    \centering
    \includegraphics[width = 0.4\textwidth]{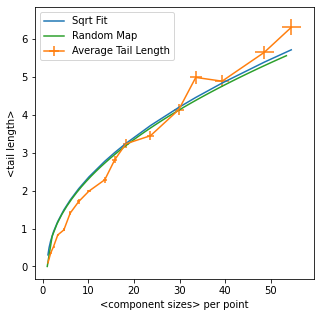}
    \caption{The orange line is the average tail length given the average component size for graphs made using local maps. The blue line is a square root best fit. The green line is for the random map graphs \cite{Flajolet:1990fk}. The behavior of tails for the local map and random map seems to be very similar.}
    \label{fig:tail_vs_component}
\end{figure}

\begin{figure}
    \centering
    \includegraphics[width = 0.4\textwidth]{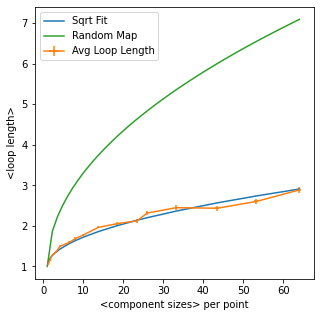}
    \caption{The orange line is the average loop length given the average component size for graphs made using local maps. The blue line is a square root fit. The green line is the average loop length for  a given average component size in random maps. The local map greatly diminishes the loop size compared to the random map, though the general shape of the graph stays the same.}
    \label{fig:loop_vs_component}
\end{figure}
\subsection{Average tail and loop lengths}

Fig. \ref{fig:loop_and_tail_vs_n} compares average tail  and loop lengths, indicating a sublinear scaling with $m$. When the average component size $C$ is used as  the independent variable, there is a close correspondence between tail lengths of our local maps and that of random maps \cite{Flajolet:1990fk}, as shown in Fig. \ref{fig:tail_vs_component}.   Remarkably, our measured average for the local maps is equal within our uncertainty to the corresponding average for random maps.  For the  random maps, this average scales as $\sqrt{C}$ \cite{Flajolet:1990fk}; this suggests that local maps have the same scaling.  As for the loops, they also show a close correspondence as shown in Fig. \ref{fig:loop_vs_component}. However, here the loop length is 2-3 times smaller than that of the random maps.  This implies that the asymptotic ratio of loop to tail length is substantially smaller in local \vs random maps, but only by a finite factor. 
\begin{figure}
    \includegraphics[width=0.4\textwidth]{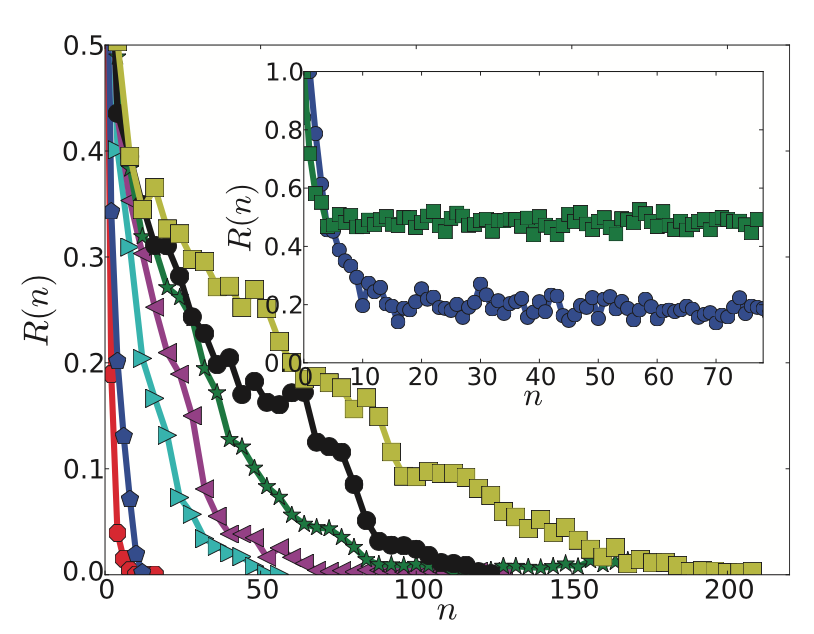}
    \caption{Graphs of energy differences $R(n)$ from Ref. \cite{Regev:2015fk}(courtesy of the authors)  versus cycle number $n$. Curves correspond to strain amplitudes  $\Gamma =$ 0.06, 0.07, 0.075, 0.085, 0.088, 0.09, 0.093, 0.095 from left to right. Inset shows graphs from strain amplitudes $\Gamma =$ 0.12 (blue) and $\Gamma = $ 0.15 (green), for which there was never a limiting cycle achieved.  See also Ref. \cite{Lavrentovich:2017rz} for similar findings.}
    \label{fig:RegevTailLengths}
\end{figure}
\begin{figure}
    \includegraphics[width=0.4\textwidth]{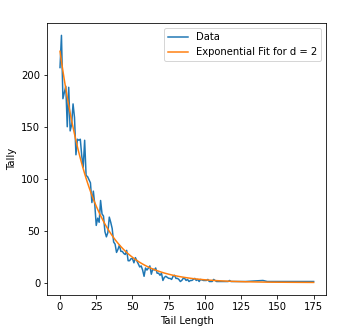}
    \caption{Tail length frequency count for 5000 generated local maps. The graph is well-fitted with an exponential decay function.}
    \label{fig:mapTailDistribution}
\end{figure}

\subsection{Tail length distribution}\label{sec:distribution}

In numerical annealing experiments, the activity during a cycle decreases to zero as the cycles proceed.   Regev \etal \cite{Regev:2015fk} monitored this decrease by measuring the net change of potential energy in a shear cycle. This $R(n)$ decreases approximately exponentially with increasing cycle number $n$, with a slower decay at larger shear amplitudes, as shown in Fig. \ref{fig:RegevTailLengths}. 

In our map picture, a large sample contains many autonomous regions each traversing its tail to arrive at a loop.  The number of active regions producing displacements at the $n$th cycle is thus the number of tails of length greater than $n$.  
Fig. \ref{fig:mapTailDistribution} shows that our tail-length distribution is also consistent with exponential decay.  The width of the exponential is determined by the average tail length, reported above. Thus the map model is qualitatively consistent with the  $R(n)$ of granular experiments.

\section{Discussion}\label{sec:Discussion}
The preliminary exploration described above suggests that the convergent annealing of granular packs may be usefully treated as an instance of dynamics on discrete sets.  The best-known examples of this dynamics are cellular automatons \cite{wolfram1984cellular}, wherein the states are spatial arrays of binary degrees of freedom.  In this work, we explore the predictive consequences of this type of dynamics for granular annealing.  Specifically, we aim to understand the collapse of the large set of initial states into a much smaller set of final states, steerable by the annealing protocol.  

The primitive constraint of locality studied above allows an improved correspondence between the discrete map dynamics and the observed granular annealing dynamics of Refs. \cite{Regev:2015fk, Lavrentovich:2017rz}. It creates convergence in a number of cycles that is independent of system size, unlike unrestricted maps.  It also resembles the progression toward the final cycle seen in these annealing simulations.  However it fails to account for a striking feature of annealing: the dominance of final states with the same period as the shear cycle, \ie loops with a single node. Further, our model shows no counterpart of the observed transition from convergence to nonconvergence, \ie the transition shown in Fig. \ref{fig:simulation_cycles}b.  These indicate that our primitive implementation of locality does not capture the essential elements of the annealing process.  

Below we consider further qualitative aspects of map properties that might account for these unexplained features.

\subsection{Implementing locality}\label{sec:implementingLocality}
Our rationale for considering locality was the observed locality of slip events near specific positions in the pack.  Yet in our model, the objects with definite positions were not the events.  Rather, they were the nodes, \ie the configurations of the system. This notion of locality has no clear justification from the granular behavior.  The direct counterpart of the granular locality would be the links of the map. Thus it would be appropriate to assign the {\em edges} of the map graph to definite positions, and to restrict the maps based on this locality.  We did not see a way to do this.

Even granting the notion of localized nodes, there is no obvious reason that the distance measure should be that of a two-dimensional space, as we have assumed.  We explored the effect of placing our nodes in three or four dimensions.  The resulting maps could not be made large enough to show any useful scaling behavior.

\subsection{Relation to hysteron-based models}
The minimalist approach used here complements hysteron models that restrict the nature of the elementary slip events.  Though we do not specify the events, hysteron events are consistent with our locality picture. The physical slip events certainly influence each other as explored \eg in \cite{Edwards:1985dq,Keim:2021aa,Lindeman:2021aa}.  We have given no account of this influence in our model.  This mutual influence could well be the key to explaining the transition to chaos and to characterizing the final states. This would imply that our model misses the point.

\section*{Conclusion}
The rich and controllable self-organization seen in cycling annealing arises in part from the qualitative features of discrete, deterministic dynamics. The principles underlying such self organization surely apply much more broadly.  The present exploratory study suggests a pathway towards identifying these principles.

\backmatter

\bmhead{Acknowledgments}
TW is grateful to Sidney Nagel and the participants at the 2022 Annual Meeting of the Simons Collaboration on Cracking the Glass transition \cite{Nagel:2022aa} for helpful discussions early in this work.  AM acknowledges partial support from the University of Chicago Materials Research Science and Engineering Center, which is funded by the National Science Foundation under award number DMR-2011854.

\bmhead{Declarations}
\begin{itemize}
\item  The authors have no relevant financial or non-financial interests to disclose.
\item The authors contributed equally to the work. AM co-developed the research plan, devised and executed all the numerical work, identified relevant references, prepared the figures and co-wrote the article.
 TW formulated the model to be studied, suggested variations and quantities to measure, and co-wrote the article. 	
 \item Data availability statement:

 The data supporting the findings of this study are available within the paper, and in the archive at \cite{Movsheva2023Paper-archive}
\end{itemize}
\appendix
\section*{Appendix}
Here we describe further details used in generating and measuring our maps.
\section{Finding components with no leaves}\label{app:noLeaves}
To complete the component-finding procedure described in Sec. \ref{sec:componentStatistics} we describe our procedure for identifying components with no leaves, consisting entirely of a loop. To find these leafless components, we have to check if any nodes in the full list of nodes haven’t been identified. Any found unidentified node in this list would be a part of a loop full of other unidentified nodes. As before, we give this node a unique identifier, and using its pointer we iterate through the loop, giving each node the same identifier as the first node in the loop. To make sure not to get stuck in this loop when following the pointers, we must check that the next node hasn’t been already identified. If it has, it means we have made a full circle in the loop. After this, we continue going down the list of all the nodes checking for unidentified nodes. Since the nodes are objects, the list gets dynamically updated as we follow a node’s pointer and update any nodes in its path; thus we don’t have to update the list itself.

\section{Finding nodes of large components}\label{app:largeComponents}
If a map is confined to a nonperiodic box, a graph that extends to the edge, in general, has missing nodes that would have been included if the initial box had been enlarged.  Here we describe our procedure for eliminating this distortion in our local maps.

We generated nodes with coordinates within a limited area for practicality, so we have to be careful of what happens on the edges of this area. When we build an iterative directed map (starting from a random node and composed only of tail and loop) its nodes could happen to be next to the boundary of the area populated by nodes. Without any alterations, the directed graph’s node proximate to the boundary would be prohibited from sampling potential image nodes on the other side of the boundary and be forced to select its image confined to the area where the nodes have been pre-generated. We are interested in the loop and tail size statistics for such graphs so this effect from the boundary could bias these statistics by prematurely forcing the directed graph to go back to the area where it has already been and have a higher chance of closing and forming a loop. To fix this, whenever during the building of a graph a node comes close enough to the boundary, we generate nodes with coordinates in that direction at the fixed density to match the previously generated node coordinates.

In order to find components unhindered by the boundaries of this space, we generate nodes with coordinates filling a large enough area (a square) at a fixed average density. Once all the image nodes for all the nodes are found, the component that is at the center of the area should not go close enough to the boundary, lest the nodes could sample outside the boundary (the nodes are within $\dmax$ of the boundary). If the component’s nodes do come too close to the boundary, to remedy this we just generated nodes for an even larger area and rebuild the maps.  This precaution is so that during the building of the maps, the boundary does not deprive the component of growing beyond the set area of generated coordinates of the nodes. 

\bibliography{MoshevaWitten2023}

\end{document}